\documentclass[twocolumn]{aastex61}
\pdfoutput=3 
\usepackage{amsmath,amstext}
\usepackage[T1]{fontenc}
\usepackage{apjfonts} 
\usepackage[breaklinks]{hyperref}
\usepackage[figure,figure*]{hypcap}
\usepackage{graphicx}
\usepackage{verbatim}
\usepackage{rotating}

\usepackage{longtable}

\begin{document}

\title{PSR J2234$+$0611: A new laboratory for stellar evolution}

\author[0000-0002-7261-594X]{K.~Stovall}
\affiliation{National Radio Astronomy Observatory, 1003 Lopezville Road, Socorro, NM, 87801, USA}

\author[0000-0003-1307-9435]{P.~C.~C.~Freire}
\affiliation{Max-Planck-Institut f\"{u}r Radioastronomie, Auf dem H\"{u}gel 69, 53131 Bonn, Germany}

\author[0000-0003-4453-3776]{J. Antoniadis}
\affiliation{Max-Planck-Institut f\"{u}r Radioastronomie, Auf dem H\"{u}gel 69, 53131 Bonn, Germany}

\author[0000-0001-8640-8186]{M.~Bagchi}
\affiliation{The Institute of Mathematical Sciences, Chennai, India 600113}

\author[0000-0003-1226-0793]{J.~S.~Deneva}
\affiliation{George Mason University, resident at the Naval Research Laboratory, 4555 Overlook Ave. SW, Washington, DC 20375, USA}

\author[0000-0001-6166-9646]{N.\,Garver-Daniels}
\affiliation{Department of Physics and Astronomy, West Virginia University, P.O. Box 6315, Morgantown, WV 26506, USA}
\affiliation{Center for Gravitational Waves and Cosmology, West Virginia University, Chestnut Ridge Research Building, Morgantown, WV 26505, USA}

\author[0000-0003-0669-865X]{J.~G.~Martinez}
\affiliation{Max-Planck-Institut f\"{u}r Radioastronomie, Auf dem H\"{u}gel 69, 53131 Bonn, Germany}

\author[0000-0001-7697-7422]{M.\,A.\,McLaughlin}
\affiliation{Department of Physics and Astronomy, West Virginia University, P.O. Box 6315, Morgantown, WV 26506, USA}
\affiliation{Center for Gravitational Waves and Cosmology, West Virginia University, Chestnut Ridge Research Building, Morgantown, WV 26505, USA}

\author{Z.\,Arzoumanian}
\affiliation{Center for Research and Exploration in Space Science and Technology and X-Ray Astrophysics Laboratory, NASA Goddard Space Flight Center, Code 662, Greenbelt, MD 20771, USA}

\author[0000-0003-4046-884X]{H.\,Blumer}
\affiliation{Department of Physics and Astronomy, West Virginia University, P.O. Box 6315, Morgantown, WV 26506, USA}
\affiliation{Center for Gravitational Waves and Cosmology, West Virginia University, Chestnut Ridge Research Building, Morgantown, WV 26505, USA}

\author{P.\,R.\,Brook}
\affiliation{Department of Physics and Astronomy, West Virginia University, P.O. Box 6315, Morgantown, WV 26506, USA}
\affiliation{Center for Gravitational Waves and Cosmology, West Virginia University, Chestnut Ridge Research Building, Morgantown, WV 26505, USA}

\author[0000-0002-6039-692X]{H.\,T.\,Cromartie}
\affiliation{University of Virginia, Department of Astronomy, P.O. Box 400325, Charlottesville, VA 22904, USA}

\author[0000-0002-6664-965X]{P.\,B.\,Demorest}
\affiliation{National Radio Astronomy Observatory, 1003 Lopezville Road, Socorro, NM, 87801, USA}

\author[0000-0002-2185-1790]{M.\,E.\,DeCesar}
\affiliation{Department of Physics, Lafayette College, Easton, PA 18042, USA}

\author[0000-0001-8885-6388]{T.\,Dolch}
\affiliation{Department of Physics, Hillsdale College, 33 E. College Street, Hillsdale, Michigan 49242, USA}

\author{J.\,A.\,Ellis}
\affiliation{Infinia ML, 202 Rigsbee Avenue, Durham NC, 27701}

\author[0000-0002-2223-1235]{R.\,D.\,Ferdman}
\affiliation{School of Chemistry, University of East Anglia, Norwich, NR4 7TJ, United Kingdom}

\author{E.\,C.\,Ferrara}
\affiliation{NASA Goddard Space Flight Center, Greenbelt, MD 20771, USA}

\author[0000-0001-8384-5049]{E.\,Fonseca}
\affiliation{Department of Physics, McGill University, 3600  University St., Montreal, QC H3A 2T8, Canada}

\author{P.\,A.\,Gentile}
\affiliation{Department of Physics and Astronomy, West Virginia University, P.O. Box 6315, Morgantown, WV 26506, USA}
\affiliation{Center for Gravitational Waves and Cosmology, West Virginia University, Chestnut Ridge Research Building, Morgantown, WV 26505, USA}

\author{M.\,L.\,Jones}
\affiliation{Department of Physics and Astronomy, West Virginia University, P.O. Box 6315, Morgantown, WV 26506, USA}
\affiliation{Center for Gravitational Waves and Cosmology, West Virginia University, Chestnut Ridge Research Building, Morgantown, WV 26505, USA}

\author[0000-0003-0721-651X]{M.\,T.\,Lam}
\affiliation{Department of Physics and Astronomy, West Virginia University, P.O. Box 6315, Morgantown, WV 26506, USA}
\affiliation{Center for Gravitational Waves and Cosmology, West Virginia University, Chestnut Ridge Research Building, Morgantown, WV 26505, USA}

\author[0000-0003-1301-966X]{D.\,R.\,Lorimer}
\affiliation{Department of Physics and Astronomy, West Virginia University, P.O. Box 6315, Morgantown, WV 26506, USA}
\affiliation{Center for Gravitational Waves and Cosmology, West Virginia University, Chestnut Ridge Research Building, Morgantown, WV 26505, USA}

\author[0000-0001-5229-7430]{R.\,S.\,Lynch}
\affiliation{Green Bank Observatory, P.O. Box 2, Green Bank, WV 24944, USA}

\author[0000-0002-3616-5160]{C.\,Ng}
\affiliation{Department of Physics and Astronomy, University of British Columbia, 6224 Agricultural Road, Vancouver, BC V6T 1Z1, Canada}
\affiliation{Dunlap Institute, University of Toronto, 50 St. George St., Toronto, ON M5S 3H4, Canada}

\author[0000-0002-6709-2566]{D.\,J.\,Nice}
\affiliation{Department of Physics, Lafayette College, Easton, PA 18042, USA}

\author[0000-0001-5465-2889]{T.\,T.\,Pennucci}
\affiliation{Hungarian Academy of Sciences MTA-ELTE ``Extragalatic Astrophysics'' Research Group, Institute of Physics, E\"{o}tv\"{o}s Lor\'{a}nd University, P\'{a}zm\'{a}ny P. s. 1/A, Budapest 1117, Hungary}

\author[0000-0001-5799-9714]{S.\,M.\,Ransom}
\affiliation{National Radio Astronomy Observatory, 520 Edgemont Road, Charlottesville, VA 22903, USA}

\author[0000-0002-6730-3298]{R.\,Spiewak}
\affiliation{Centre for Astrophysics and Supercomputing, Swinburne University of Technology, P.O. Box 218, Hawthorn, Victoria 3122, Australia}

\author[0000-0001-9784-8670]{I.\,H.\,Stairs}
\affiliation{Department of Physics and Astronomy, University of British Columbia, 6224 Agricultural Road, Vancouver, BC V6T 1Z1, Canada}

\author{J.\,K.\,Swiggum}
\affiliation{Center for Gravitation, Cosmology and Astrophysics, Department of Physics, University of Wisconsin-Milwaukee, P.O. Box 413, Milwaukee, WI 53201, USA}

\author[0000-0003-4700-9072]{S.\,J.\,Vigeland}
\affiliation{Center for Gravitation, Cosmology and Astrophysics, Department of Physics, University of Wisconsin-Milwaukee, P.O. Box 413, Milwaukee, WI 53201, USA}

\author{W.\,W.\,Zhu}
\affiliation{National Astronomical Observatories, Chinese Academy of Science, 20A Datun Road, Chaoyang District, Beijing 100012, China}

\correspondingauthor{K.~Stovall}
\email{kstovall@nrao.edu}


\begin{abstract}
We report the timing results for PSR J2234$+$0611, a 3.6-ms pulsar in a 
32-day, eccentric ($e\,=\,0.13$) orbit with a helium white dwarf. The 
precise timing and eccentric nature of the orbit allow measurements of
an unusual number of parameters: a) a precise proper motion of 27.10(3) 
$\mathrm{mas\;yr^{-1}}$ and a parallax of 1.05(4) mas resulting in a 
pulsar distance of 0.95(4) kpc; enabling an estimate of the transverse 
velocity, 123(5) $\mathrm{km\;s^{-1}}$. Together with previously 
published spectroscopic measurements of the systemic radial velocity, 
this allows a 3-D determination of the system's velocity; b) precise 
measurements of the rate of advance of periastron yields a total system 
mass of $1.6518^{+0.0033}_{-0.0035}$ M$_\odot$; c) a Shapiro delay 
measurement, $h_3\,=\,82\pm14$ ns despite the orbital inclination not 
being near 90$^\circ$; combined with the measurement of the total mass 
yields a pulsar mass of $1.353^{+0.014}_{-0.017}\,\mathrm{M_{\odot}}$ 
and a companion mass of $0.298^{+0.015}_{-0.012}\mathrm{M_\odot}$; d) 
we measure precisely the secular variation of the projected semi-major 
axis and detect significant annual orbital parallax; together these 
allow a determination of the 3-D orbital geometry of the system, 
including an unambiguous orbital inclination 
($i\,=\,138.7^{+2.5}_{-2.2}\deg$) and a position angle for the line 
of nodes ($\Omega\,= \,44^{+5}_{-4}\deg$). We discuss the component 
masses to investigate the hypotheses previously advanced to explain the
origin of eccentric MSPs. The unprecedented determination of the 3-D 
position, motion and orbital orientation of the system, plus the precise 
pulsar and WD masses and the latter's optical detection make this system 
an unique test of our understanding of white dwarfs and their atmospheres.
\end{abstract}

\section{Introduction}\label{sec:intro}
Millisecond pulsars (MSPs) are a population of pulsars with much faster spin rates and significantly smaller
spin-down rates than that of the ``normal'' pulsars. They are believed to be formed through a process in which
a neutron star (NS) goes through a long period of accretion from a companion star. This mass transfer process
circularizes the orbit and results
in the neutron star spinning faster and a reduction in the neutron star's magnetic field. If the companion
is a low-mass star, then the system is seen during accretion as a low-mass X-ray
binary~\citep[LMXB;][]{1982Natur.300..728A,1982CSci...51.1096R}. The tidal circularization for these
systems results invariably in orbits with very low eccentricities.
The result of the evolution of a LMXB is a MSP orbited by a helium white dwarf (He WD).
A fundamental expectation of this process is that the orbit of a MSP - He WD should have a 
very low eccentricity \citep{1992RSPTA.341...39P}, since the formation of the companion He WD is not
associated with violent events, like supernova explosions. This is
confirmed by the very small eccentricities measured for the vast majority of MSPs with He WD companions.

\begin{figure*}
\begin{center}
\includegraphics[width=0.8\textwidth]{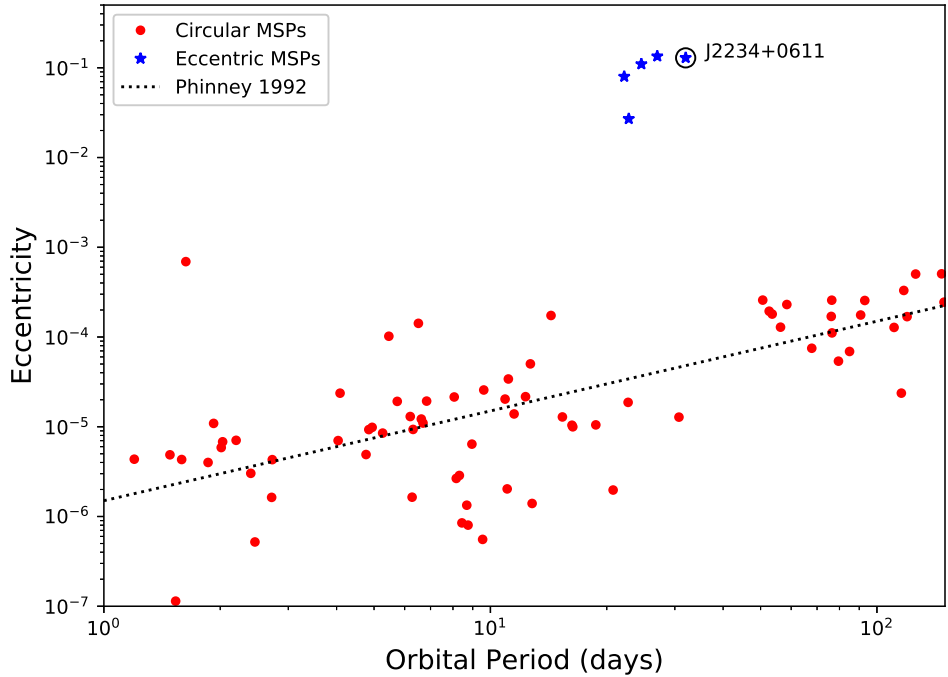}
\caption{Eccentricity ($e$) versus orbital period ($P_{\rm b}$) for recycled pulsars
with low-mass ($< 0.6 \, \mathrm{M_{\odot}}$) companions outside globular clusters. The eccentric MSPs, blue stars,
stand out from the general MSP population, red circles. For the latter, the orbital eccentricities are
small and generally follow the evolution predicted by \cite{1992RSPTA.341...39P}, shown by the black dotted line.
Note that there is an under-density of circular MSP systems within the orbital period range
where the eccentric MSPs are found and going to larger orbital periods, as first
noted by~\citep{1995PhDT.........5C}.}
\label{fig:ecc_pb}
\end{center}
\end{figure*}

In recent years, a small set of systems that are inconsistent with the
typical formation scenario have been discovered in the Galactic field: PSRs~J0955$-$6150 
\citep{2015ApJ...810...85C}, J1618$-$3921~\citep{2001ApJ...553..801E,2018A&A...612A..78O},
J1946$+$3417 \citep{2013MNRAS.435.2234B}, J1950$+$2414 \citep{2015ApJ...806..140K}  and
J2234$+$0611 \citep{2013ApJ...775...51D}; the latter will be the focus of this work.
 All have orbital eccentricities
in the range 0.027 - 0.14 and small mass ($\sim \, 0.3 \, M_{\odot}$) companions.
Additionally, the orbital periods for these systems are quite similar ($P_b\sim 22\,-\,32$ d,
see Figure~\ref{fig:ecc_pb}).

The first known MSP with an eccentric
orbit in the Galactic field, PSR~J1903+0327 \citep{2008Sci...320.1309C} (with an orbital period of 95 d and
orbital eccentricity of 0.43, the companion is a 1.03 $\mathrm{M_{\odot}}$ main sequence star),
is thought to have formed in the chaotic disruption of a triple system \citep{2011MNRAS.412.2763F}.
This is not a likely explanation for the former systems given the similarity of their
orbital parameters. A number of hypotheses for their formation have been advanced, including
rotationally delayed accretion induced
collapse~\citep{2014MNRAS.438L..86F}, a phase transition inside the MSP that results
in the formation of a strange star core \citep{2015ApJ...807...41J} and eccentricity pumping
via interaction with a circumbinary disk~\citep{2016ApJ...830...36A}.

In wide, circular MSP systems, the only relativistic parameters that
can be measured are the `range' ($r$) and `shape' ($s$) parameters of a Shapiro delay. Such measurements
are only possible for systems with high orbital inclinations and where the pulsar has high timing precision,
or the companion is massive. The result is that only four systems have NS mass measurements better
than 5\% from Shapiro delay alone (PSR~J2222$-$0137, \citealt{2017ApJ...844..128C}, PSRs J1909$-$3744,~J1614$-$2230 and J1713$+$0747, \citealt{2018ApJS..235...37A}).
If the wide MSP binary is eccentric, then we can also measure the advance
of periastron ($\dot{\omega}$), which gives a measurement of the total system mass ($M_{\rm tot}$).
This, together with even a poorly determined Shapiro delay, allows the measurement of precise MSP masses
\citep{2011MNRAS.412.2763F,2012ApJ...745..109L,2017MNRAS.465.1711B}, but this is a relatively rare occurrence
since eccentric MSP systems are rare. Using this technique, the 
masses
of two of the eccentric MSPs, PSRs~J1946+3417 and J1950+2414, have already been measured precisely by \cite{2017MNRAS.465.1711B} and Zhu et al. (in preparation);
the pulsar masses are $1.828(22)\, \rm M_{\odot}$ and $1.495(24)\, M_{\odot}$ respectively
and the companion masses are $0.2556(19) \, \rm M_{\odot}$ and $0.280_{-0.004}^{+0.006}\, M_{\odot}$
respectively.

In this paper, we present a study of PSR J2234+0611, an eccentric MSP system for which the precise timing has,
as in the case of PSR~J1946+3417 and J1950+2414, enabled precise mass measurements for both the pulsar and
its companion. In Section~\ref{sec:obs}, we detail the detection and follow-up timing observations.
In Section~\ref{sec:results}, we describe the phenomenological development of the
timing model, enumerating the different
orbital effects that are detectable in this system and present some initial results. In 
Section~\ref{sec:bayesian}, we extend on the preliminary results using Bayesian methods
to determine the masses and orbital orientation of the system in a self-consistent way, 
assuming the validity of general relativity.
In Section~\ref{sec:discussion}, we discuss the implications of our findings.
In Section~\ref{sec:conc}, we summarize our conclusions for this system.

Some of these results have already been presented preliminarily by \cite{2016ApJ...830...36A},
who confirmed, based on the timing position of the system, that the companion
is a He WD. From the spectroscopy of the WD, they placed limits on the systemic radial velocity of the system,
$\mathrm{V_r\approx -20(34)\;km\;s^{-1}}$. They used this, together with our preliminary timing values
for the proper motion and the distance, to study the system's 3-D motion in the Galaxy.

\section{Observations and Analysis}\label{sec:obs}

\subsection{Discovery and observations}

PSR J2234$+$0611 was discovered in the Arecibo Observatory 327 MHz Drift Scan Survey in December
2012~\citep{2013ApJ...775...51D}. After discovery of the pulsar, 
initial follow-up observations were performed, also with the Arecibo 305-m telescope,
using the ``L-wide'' receiver at a center frequency
of 1.5 GHz and recorded with the Puerto Rican Ultimate Pulsar
Processing Instrument (PUPPI) in search mode, allowing for offline
folding of each observation to get the observed period
at each epoch. The preliminary orbital parameters resulting
from these observations were
already reported in \cite{2013ApJ...775...51D}.

\begin{figure*}
\begin{center}
\includegraphics[width=0.7\columnwidth]{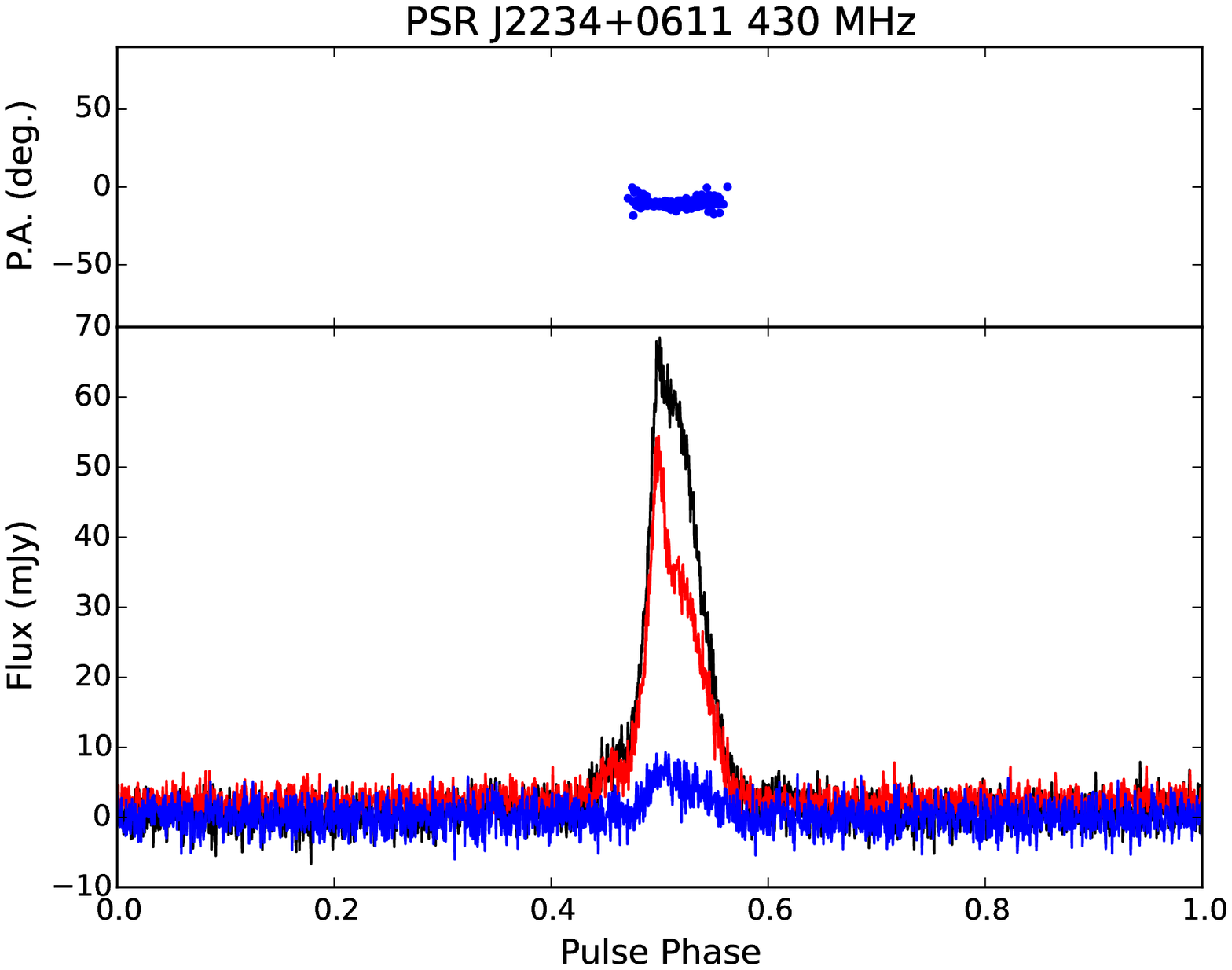}
\includegraphics[width=0.7\columnwidth]{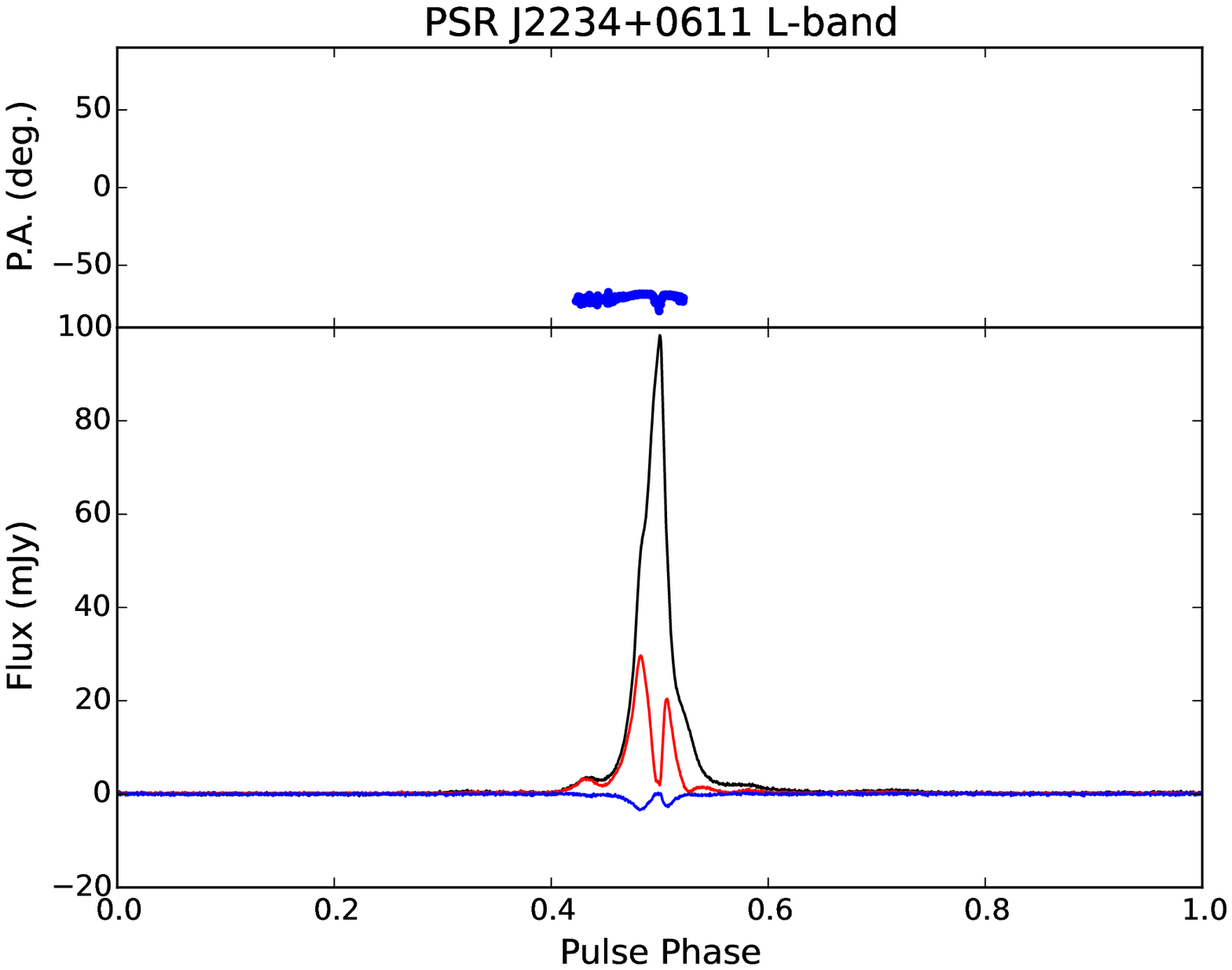}
\caption{Polarimetric profiles for PSR J2234+0611 at 430 MHz (left) and 1.5 GHz (right) from the Arecibo Observatory using the PUPPI backend with bandwidths of 20 and 650 MHz, respectively. These profiles were taken from individual high signal-to-noise ratio detections. The top panels show the polarization angle versus pulse phase. The bottom panels show the total intensity (black), linear polarization (red), and circular polarization (blue) versus pulse phase. The profiles have been polarization and flux calibrated using the methods described in ~\cite{2018ApJS..235...37A}. These profiles have not been corrected for rotation measure, as the value measured from these observations is consistent with 0 $\mathrm{rad\;m^{-2}}$. Additional analysis of the polarization properties for PSR J2234$+$0611 has been presented in~\cite{2018ApJ...862...47G}.}
\label{fig:profile}
\end{center}
\end{figure*}

We then folded the data using the new orbit and began to
refine the timing solution by generating pulse times-of-arrival (ToAs) and performing pulsar timing analysis using
{\tt tempo}\footnote{\url{http://tempo.sourceforge.net/}}. Subsequent data was recorded using PUPPI in coherent dedispersion and online folding mode. Figure~\ref{fig:profile} shows the profile
for PSR J2234+0611 at 430 and 1.5 GHz from roughly 30-minute duration coherent fold mode observations.

PSR J2234+0611 was immediately found to have excellent timing precision and therefore was added to
the pulsar timing arrays (PTAs) efforts to detect low frequency gravitational waves,
in particular to the North American Nanohertz 
Observatory for Gravitational Waves \citep[NANOGrav,][]{2013ApJ...762...94D} PTA.
Observations of the pulsar have continued under that project, using the Arecibo 305-m radio telescope
with the ``L-wide'' receiver (with frequency coverage between 1130 and 1730 MHz) and the 
430 MHz receiver with a cadence of about 3 weeks. For both types of observations,
the PUPPI back-end was used, with coherent dedispersion and folding
mode, as for other PTA pulsars; these observations are described in detail by \cite{2018ApJS..235...37A} but extend later in time than
the data presented in that paper.
Current timing solution parameters from data spanning 5 years are given in Tables~\ref{table:timsol}
and \ref{tab:orbital_solutions}.

\begin{figure*}
\begin{center}
\includegraphics[width=0.9\textwidth, angle=0]{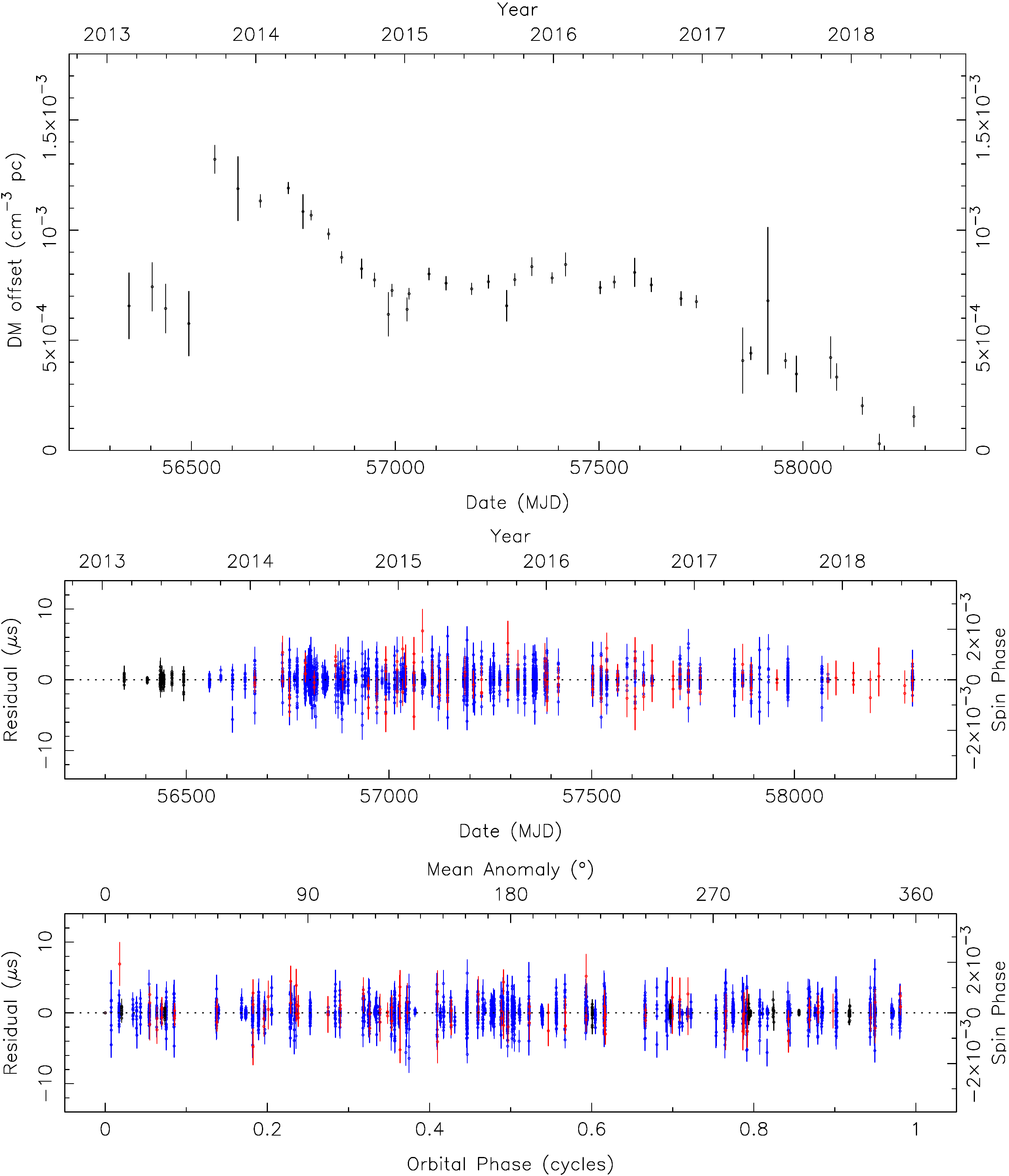}
\end{center}
\caption{Five years of high-precision timing data for PSR~J2234+0611. 
{\em Top}: Dispersion measure offsets relative
to the reference DM (10.778 cm$^{-3}\,$pc) as a function of date.
{\em Middle}: ToA residuals for the DDGR ephemeris in Table~\ref{table:timsol}
as a function of date, and {\em Bottom}: ToA residuals as a function of the
orbital phase. The residual 1-$\sigma$ uncertainties are indicated by
vertical error bars. Black indicates the data from the initial incoherent
observations at 1.5 GHz, blue data from the coherent observations at 1.5 GHz and 
red the coherent observations at 430 MHz. There is a jump in the measured DM 
offsets between the incoherent observations and the coherent observations due to
covariances between the DM offsets and a constant offset between the two
datasets.
}
\label{fig:residuals}
\end{figure*}

\subsection{Timing analysis}

The timing analysis of the PUPPI data is similar to that described by \cite{2018ApJS..235...37A}.
The ToAs 
are derived from the integrated pulse profiles
using the standard {\tt PSRCHIVE} routines. The ToA analysis is made using {\tt tempo}.
To convert the telescope ToAs (corrected to the International Bureau of Weights and Measures
version of Terrestrial Time, TT) to the Solar System barycentre, we used the
Jet Propulsion Laboratory's DE436 solar system ephemeris; the
resulting timing parameters are presented in Barycentric Dynamical Time
(TDB). We used the same method used by NANOGrav to estimate variations of the
dispersion measure (DM), but with the ToAs grouped in intervals of
32 days (the orbital period), instead of 6 days as is the norm for the NANOGrav pulsars.
DM values are reported as offsets relative to an arbitrary fiducial
value of 10.778 pc cm$^{-3}$.

We used three orbital models to analyze the data, all based on the
description of \cite{1985AIHS...43..107D,1986AIHS...44..263D}.
The first is the ``DDGR'' model, which assumes the validity of
general relativity (GR) and where we fit directly for the
total mass of the system ($M_{\rm tot}$) and the companion mass ($M_c$).
The second model is basically the theory-independent DD model, but with the
orthometric parameterization
of the Shapiro delay described by \cite{2010MNRAS.409..199F}; this
was implemented in {\tt tempo} by \cite{2016ApJ...829...55W}, where it is designated as
the ``DDFWHE" orbital model. The third model is again based on the
DD model but takes into account the kinematic effects described by
\cite{1995ApJ...439L...5K,1996ApJ...467L..93K}; this was implemented in {\tt tempo2}
by \cite{2006MNRAS.372.1549E}, where it is designated as the ``T2'' model; it was implemented in {\tt tempo}
by one of us (IHS), where it is designated as the
``DDK'' model.

The reason for the usage of these three orbital models is
that, as we will show, no single model alone fully captures all the constraints
on the masses and orbital orientation of this system.
In the DDFWHE and DDK solutions,
we used the Einstein delay calculated in the DDGR solution;
the reason for this is because it cannot be determined independently
with our data, and because it is strongly correlated with $\dot{x}$
(see Ridolfi et al. 2018, in preparation). Furthermore, the
orthometric ratio of the Shapiro delay ($\varsigma$)
in the DDFWHE solution and the orbital inclination ($i$) in the
DDK solution are derived from the $s \, \equiv \, \sin i$ parameter
calculated by the DDGR solution; the reason being the extremely
small signature of the Shapiro delay.
In Section~\ref{sec:results}, we discuss the significance of these parameters.

\subsection{Flux Measurements}
As part of the NANOGrav data analysis procedures, the data have been flux and
polarization calibrated, allowing straightforward measurements of the polarization
profile (Figure~\ref{fig:profile}) and mean flux density. We have taken flux density values
from a preliminary analysis of the upcoming 12.5 year data release~(Arzoumanian et al.,
in prep.). The data in this preliminary release was polarization and flux calibrated
using the same methods as the NANOGrav 9-year data release~\citep{2015ApJ...813...65T}.
However, the observed flux density for PSR J2234$+$0611 varies over a fairly wide
range due to scintillation by the interstellar medium. Using {\tt psrflux} from the {\tt PSRCHIVE} pulsar
suite, we calculated the mean value from 43 observations
at 430 MHz, ranging from 0.2 to 5.3 mJy and 50 observations at 1.5 GHz, ranging from 0.03 to
3.3 mJy to get an estimate for the mean flux density at these frequencies. The resulting
mean values and spectral index are given in Table~\ref{table:timsol}.

\begin{table}
\begin{center}{\scriptsize
\caption{Non-binary parameters for PSR~J2234+0611\label{table:timsol}}
\setlength\extrarowheight{1pt}
\begin{tabular}{l c}
\hline
  \hline
  \multicolumn{2}{l}{Observation and data reduction parameters}\\
  \hline
  Reference Epoch (MJD) \dotfill & 56794.093186 \\
  Span of timing data (MJD) \dotfill & 56347 - 58291 \\
  Number of ToAs \dotfill & 5882 \\
  Solar wind parameter, $n_0$ (cm$^{-3}$) \dotfill & 6 \\
  Overall individual ToA RMS residual ($\mu s$) \dotfill & 0.58 \\
  RMS residual for incoherent L-band ($\mu s$) \dotfill & 0.35 \\
  RMS residual for coherent 430 MHz ($\mu s$) \dotfill & 1.51 \\
  RMS residual for coherent L-band ($\mu s$) \dotfill & 0.59 \\
  $\chi^2$ \dotfill & 5891.76 \\
  Reduced $\chi^2$ \dotfill & 1.013  \\
  \hline
  \multicolumn{2}{l}{Spectral parameters}\\
  \hline
  Mean flux density at 430 MHz, $S_{430}$ (mJy) \dotfill & 1.3\\
  Mean flux density at 1400 MHz, $S_{1400}$ (mJy) \dotfill & 0.6 \\
  Spectral Index, $\alpha$ \dotfill & $-0.7$\\
  \hline
  \multicolumn{2}{l}{Astrometric and spin parameters}\\
  \hline
  Right ascension, $\alpha$ (J2000) \dotfill & 22:34:23.073090(2) \\    
  Declination, $\delta$ (J2000) \dotfill & 06:11:28.68633(7) \\      
  Proper motion in $\alpha$, $\mu_\alpha$ ($\mathrm{mas\;yr^{-1}}$) \dotfill & 25.30(2)\\     
  Proper motion in $\delta$, $\mu_\delta$ ($\mathrm{mas\;yr^{-1}}$) \dotfill & 9.71(5) \\  
  Parallax, $\varpi$ (mas) \dotfill & 1.03(4) \\
  Spin frequency, $\nu$ (Hz) \dotfill & 279.5965821510426(5) \\
  Spin frequency derivative, $\dot{\nu}$ ($10^{-16}\, \rm Hz \, s^{-1}$) \dotfill & $-$9.3920(1) \\
  Dispersion measure, DM ($\rm pc\,  cm^{-3}$) \dotfill & 10.778 \\
  \hline
  \multicolumn{2}{l}{Derived parameters}\\
  \hline
  Galactic longitude, $l$ \dotfill & +72.99 \\
  Galactic latitude, $b$ \dotfill & $-$43.01 \\
  Magnitude of proper motion, $\mu$ ($\rm mas \, yr^{-1}$) \dotfill & 27.10(2) \\
  Position angle of proper motion, $\Theta_{\mu}$ ($\deg$, J2000) \dotfill & 69.0(1) \\
  Position angle of proper motion, $\Theta_{\mu}$ ($\deg$, Galactic) \dotfill & 111.5(1)\\  
  DM-derived distance, $d_1$ (kpc) \dotfill & 0.68 \\
  DM-derived distance, $d_2$ (kpc) \dotfill & 0.86 \\
  Parallax-derived distance, $d$ (kpc) \dotfill & 0.97(4) \\
  Galactic height, $z$ (kpc) \dotfill & -0.651(26) \\
  Transverse velocity, $v_{\rm T}$ ($\rm km \, s^{-1}$) \dotfill & 123(5) \\
  Spin period, $P$ (ms) \dotfill & 3.576581631673107(6) \\
  Spin period derivative, $\dot{P}$ ($10^{-20}$~s s$^{-1}$) \dotfill  & 1.20142(1) \\
  Intrinsic spin period derivative, $\dot{P}_{\rm int}$ ($10^{-20}$~s s$^{-1}$) \dotfill  &  $0.647^{+0.023}_{-0.025}$ \\
  Surface magnetic flux density, $B_0$ ($10^{8}$ Gauss) \dotfill & 1.5 \\
  Characteristic age, $\tau_c$ (Gyr) \dotfill &  8.8 \\
Spin-down power, $\dot{E}$ ($10^{33}\, \rm erg \, s^{-1}$) \dotfill & 5.6 \\
  \hline
\multicolumn{2}{l}{Notes. Timing parameters and 1-$\sigma$ uncertainties derived using {\sc tempo} in}\\
\multicolumn{2}{l}{Barycentric Dynamical Time (TDB), using the DE 421 Solar System ephemeris,}\\
\multicolumn{2}{l}{the NIST UTC time timescale and the DDGR orbital model.}\\
\multicolumn{2}{l}{$d_1$ is derived using the NE2001 \citep{2002astro.ph..7156C} Galactic model,}\\
\multicolumn{2}{l}{$d_2$ using the YMW16 \citep{2017ApJ...835...29Y} Galactic model.}\\
\multicolumn{2}{l}{Estimate of $v_{\rm T}$, $\dot{P}_{\rm int}$ and derived parameters assume distance from measured}\\
\multicolumn{2}{l}{parallax and its uncertainty.}
\end{tabular}
\vspace{-0.5cm}
}
\end{center}
\end{table}

\begin{table*}
\begin{center}{\footnotesize
\caption{Orbital parameters for PSR~J2234+0611\label{tab:orbital_solutions}}
\setlength\extrarowheight{1pt}
\begin{tabular}{l c c c c}
\hline
  \hline
  Orbital model \dotfill & DDGR & DDFWHE & DDK & DDK Bayesian grid \\
  Residual $\chi^2$ \dotfill & 5891.8 & 5891.7 & 5872.9 & \\
  Reduced $\chi^2$ \dotfill & 1.013 &1.013 & 1.010 & \\
\hline
  Orbital period, $P_{\rm b}$ (days) \dotfill & 32.001401626(8) & 32.001401627(8) & 32.001401630(8) & -\\
  Projected semi-major axis, $x$ (lt-s) \dotfill & 13.937366(5) & 13.9373664(3) & 13.9373664(3) & - \\
  Epoch of periastron, $T_0$ (MJD) \dotfill & 56794.0931866(1) & 56794.0931866(1) & 56794.0931866(1) & - \\
  Orbital eccentricity, $e$ \dotfill & 0.129274035(5) & 0.129274034(8) & 0.129274035(8) & - \\
  Longitude of periastron, $\omega$ ($^\circ$) \dotfill & 277.1673(2) & 277.167331(1) & 277.167330(1) & - \\
  Total mass, $M_{\rm tot}$ ($\rm M_{\odot}$ ) \dotfill & 1.679(3) & - & - & $1.6518^{+0.0033}_{-0.0035}$ \\
  Companion mass, $M_c$ ($\rm M_{\odot}$ ) \dotfill & 0.300(13) & - & 0.30(5) & $0.298^{+0.015}_{-0.012}$\\
  Shapiro delay $s$ \dotfill & [0.667765] & - & - & - \\
  Rate of advance of periastron, $\dot{\omega}$ ($\deg\, \rm yr^{-1}$) \dotfill & [0.0008863] & 0.0008863(10) & 0.0008766(10) & - \\
  Einstein delay, $\gamma$ (s) \dotfill & [0.000847606] & [0.000847606] & [0.000847606] & - \\
  Derivative of $P_{\rm b}$, $\dot{P}_{\rm b}$ ($10^{-12}$ s s$^{-1}$) \dotfill & 1.8(2.5)$^{\rm a}$ & 1.9(2.5) & 3.1(2.5) & - \\
  Orthometric amplitude of Shapiro delay, $h_3$ (ns) \dotfill & - & 82(14) & - & - \\
  Orthometric ratio of Shapiro delay, $\varsigma$ \dotfill & - & $0.382811^{\rm b}$ & - & - \\
  Derivative of $x$, $\dot{x}$ ($10^{-15}$ lt-s s$^{-1}$) \dotfill 	&  $-$27.8(7) & $-$27.8(7) & - & - \\
  Orbital inclination ($\deg$) \dotfill &  - & - & $138.105^{\rm b}$ & $138.7^{+2.5}_{-2.2}$  \\
  Position angle of line of nodes, $\Omega$ ($\deg$) \dotfill  &  - & - & 43.4(7) & $44^{+5}_{-4}$ \\
  \hline
  \multicolumn{4}{l}{Derived parameters}\\
  \hline
  Mass function, $f$ ($\rm M_{\odot}$ ) \dotfill & 0.002838487(3) & 0.0028384868(2) & 0.0028384867(2) & - \\
  Pulsar mass, $M_{p}$ ($\rm M_{\odot}$ ) \dotfill & 1.38(1) & - & - & $1.353^{+0.014}_{-0.017}$\\
  \hline
\multicolumn{4}{l}{Notes. Timing parameters and 1-$\sigma$ uncertainties derived using {\sc tempo}, in Barycentric Dynamical Time (TDB)}\\
\multicolumn{4}{l}{using JPL's DE421 Solar System Ephemeris and the NIST UTC timescale.}\\
\multicolumn{4}{l}{Numbers in square brackets are derived by the DDGR model. Of these, $\gamma$ is used in the DDFWHE and DDK models.}\\
\multicolumn{4}{l}{a: Fitted as an extra contribution to the (very small) relativistic $\dot{P}_{\rm b}$ in 
the DDGR solution.}\\
\multicolumn{4}{l}{b: Assumed in the model, derived from $s$ parameter in the DDGR solution.}\\
\end{tabular}
\vspace{-0.5cm}
}
\end{center}
\end{table*}

\begin{table*}
\begin{center}
\caption {Details for grid regions.}
\label{tab:grid}
\begin{tabular}{ccccccc}
Region & $\cos{i}$ & $\Omega$ & Best $\cos(i)$ & Best $\Omega$ & Best $\rm M_{tot}$ & Min $\chi^2$\\
\hline
\hline
1 & $-$0.92 to -0.52 & 34$^\circ$ to 54$^\circ$ & -0.748 & 43.8 & 1.6512 & 5872.9 \\
2 & 0.52 to 0.92  & 90.0$^\circ$ to 110.0$^\circ$ & 0.748 & 94.4 & 1.6518 & 5881.7 \\
3 & 0.52 to 0.92  & 210.0$^\circ$ to 230.0$^\circ$ & 0.716 & 220.6 & 1.7058 & 5926.0 \\
4 & $-$0.92 to -0.52  & 270.0$^\circ$ to 290.0$^\circ$ & -0.704 & 278.4 & 1.7058 & 5929.6 \\
\hline
\end{tabular}
\end{center}
\end{table*}

\section{Results}
\label{sec:results}

The timing parameters resulting from the timing models described before are given
in Tables~\ref{table:timsol} and \ref{tab:orbital_solutions}.
The spin and astrometric parameters derived from the DDGR orbital solution
are presented in Table~\ref{table:timsol}; the reason for only presenting
this solution is that these parameters are nearly
identical for the other orbital solutions.
The orbital parameters for the three solutions are presented in
Table~\ref{tab:orbital_solutions}, as well as the results from
the Bayesian analysis described in Section \ref{sec:bayesian},
which yields the most reliable parameters and uncertainties.
We have applied EFACs, a multiplication factor for the ToA
uncertainties, and EQUADs, an error term added in quadrature
to the ToA uncertainties, for each receiver and
backend configuration, and have also allowed a fit for an
arbitrary offset between the 3 types of data; 1.5 GHz incoherent, 
430 MHz coherent, and the 1.5 GHz coherent.
For the 5882 ToAs used in our analysis, we obtain a weighted residual
root mean square (rms) of 0.58 $\mu$s and a reduced $\chi^2$ of 1.013 for the
best orbital model (DDK).
The evolution of the DM with time and the ToA residuals with time
are displayed in Fig.~\ref{fig:residuals}; the residuals are also
presented as a function of the orbital phase.

\subsection{Distance and velocity}

For this pulsar, we obtain a highly significant measurement of the 
parallax, 1.05(4) mas (all uncertainties are 68.3\% confidence limits) resulting in a 
pulsar distance $d$ of 0.95(4) kpc. This distance can be compared with the 
prediction of the DM models. The NE2001 model
\citep{2002astro.ph..7156C} predicts a distance of 0.68 kpc, while the
YMW16 model \citep{2017ApJ...835...29Y} predicts a distance of 0.86 kpc.
To these estimates is generally assigned a relative uncertainty of about 20\%.
Our parallax measurement is certainly in better agreement with 
the YMW16 model.

This measurement, together with the measurement of the proper motion, allows
a relatively accurate measurement of the Heliocentric transverse velocity,
123(5) $\mathrm{km\;s^{-1}}$.
Combining this with the systemic radial velocity
of $-20(34) \, \rm km \, s^{-1}$ measured by \cite{2016ApJ...830...36A},
we obtain a 3-D heliocentric velocity of
$124^{+10}_{-5} \mathrm{km\;s^{-1}}$.
This velocity is smaller than that used in the detailed analysis of
the Galactic motion of PSR~J2234+0611 made by \cite{2016ApJ...830...36A},
mostly because they were using a preliminary value of the parallax
that yielded a larger distance, however the qualitative conclusions
obtained by \cite{2016ApJ...830...36A} remain valid:
the 3-D velocity of this system is similar to what has been observed
for other nearby recycled pulsars (e.g., \citealt{2011ApJ...743..102G}).
We will return to this topic in Section~\ref{sec:discussion}, particularly
in the discussion on the formation of the system.

\subsection{Kinematic effects: Rate of change of Doppler shift}

For any assumed distance we can estimate the magnitude of the 
kinematic effects on the variation of the Doppler shift factor ($D$)
using the simple expressions
provided by \cite{1970SvA....13..562S} for the effect of the
centrifugal acceleration (proportional to
the square of the total proper motion, $\mu$) and \cite{1991ApJ...366..501D} for the effect
of the difference in the Galactic accelerations of the pulsar's system
and the Solar System projected along the direction from the pulsar to the Earth, $a_l$:
\begin{equation}
\frac{\dot{D}}{D} \, \equiv - \frac{\mu^2 d + a_l}{c}
\end{equation}
where $c$ is the speed of light. 
In order to estimate $a_l$, we use the expressions presented by \cite{2009MNRAS.400..805L},
where the equation for the vertical acceleration should be valid to a Galactic height
of $\sim \, \pm 1.5$ kpc (the Galactic height of PSR~J2234+0611 is $-$0.651(26) kpc).
In those expressions we use the distance to the centre of the Galaxy measured by
the GRAVITY experiment \citep{2018A&A...615L..15G}, $r_0 \, = \, 8.122(31)$ kpc
and a revised value for the rotational velocity of the Galaxy derived using the
latter $r_0$ \citep{2018arXiv180809435M}, $v_{\rm Gal} \, = \, 233.3$ km s$^{-1}$.
We obtain $a_l / c\, = \, -1.53 \, \times\, 10^{-19}\, \rm s^{-1}$
(for a comparison, we can use the Galactic model presented by \cite{2017MNRAS.465...76M} to obtain
$a_l / c\, = \, -1.76\, \times\, 10^{-19}\, \rm s^{-1}$, which is a similar number).
For the proper motion term we obtain $\mu^2 d / c \, = \, 1.702 \, \times \, 10^{-18}\, \rm s^{-1}$,
an order of magnitude larger.
Adding both terms, we obtain  $\dot{D}/D \, = \, - 1.550\, \times\, 10^{-18}\, \rm s^{-1}$. 

The  contribution of this effect to the spin period derivative is given by
$\dot{P}_{\rm kin}\, = \, - P\,  \dot{D}/D \, = \, 5.54_{-0.25}^{+0.23} \, \times \, 10^{-21}\, \rm s \, s^{-1}$.
Subtracting this from the observed $\dot{P}$ in Table~\ref{table:timsol}
we obtain the intrinsic spin period derivative
($\dot{P}_{\rm int} \, = \, 6.47_{-0.25}^{+0.23} \, \times \, 10^{-21}\, \rm s \, s^{-1}$),
which is about half of the observed $\dot{P}$.
From this and the spin period $P$, we derive 
a surface magnetic flux density $B_0\, \simeq \, 1.5 \, \times \, 10^8 \, \rm G$,
the rate of loss of rotational energy
$\dot{E}\, \simeq \, 5.6\, \times 10^{33} \, \rm erg \, s^{-1}$
and a characteristic age $\tau_c\, \simeq \, 8.8 \, \rm \, Gyr$
using the standard equations summarized by \cite{2004hpa..book.....L}.
The cooling age for the WD companion is 1.5 Gyr, which
according to \cite{2016ApJ...830...36A} is comparable to the age of the system.
This is compatible with $\tau_c$ since the latter represents
an upper limit for the age that assumes that the initial spin period
$P_{\rm init}$ was much smaller than the currently observed $P$.
Assuming a $n\, = \, 3$ braking index and an age of 1.5 Gyr,
we obtain $P_{\rm init} \, = \, 3.25 \, \rm ms$.

\subsection{Post-Keplerian effects. I. Orbital period derivative}

This rate of change of the Doppler shift factor will also be a dominant
contributor to the observed variation of the orbital period, $\dot{P}_{\rm b, obs}$.
According to \cite{2004hpa..book.....L}:
\begin{equation}
\label{eq:pbdot}
\left( \frac{\dot{P}_{\rm b}}{P_{\rm b}} \right)^{\rm obs} \, = \,
- \, \frac{\dot{D}}{D}\, +\,
\left( \frac{\dot{P}_{\rm b}}{P_{\rm b}} \right)^{\rm GW}\,  +\,
\left( \frac{\dot{P}_{\rm b}}{P_{\rm b}} \right)^{\rm \dot{m}}\, +\,
\left( \frac{\dot{P}_{\rm b}}{P_{\rm b}} \right)^{\rm T},
\end{equation}
the first term, the kinematic contribution to $\dot{P}_{\rm b, obs}$ is given by
$\dot{P}_{\rm b, kin}\, = \, - P_{\rm b}\, \dot{D}/D \, = \, 4.28^{+0.19}_{-0.18} \, \times \, 10^{-12} \, \rm \,
s^{-1}$. The second term in eq.~\ref{eq:pbdot} is due to loss of orbital energy caused by the emission of 
gravitational waves. For PSR~J2234+0611, this term is, assuming the validity of GR, given by
$\dot{P}_{\rm b, GR} \, = \,  2.62 \, \times \, 10^{-17} \, \rm s \, s^{-1}$
(this is the estimate provided by the DDGR model for the masses derived by that model). This is
about 5 orders of magnitude smaller than $\dot{P}_{\rm b, obs}$ and 
its uncertainty. The third term is caused by
radiative mass loss from the system. Assuming that this is dominated by the loss of rotational energy for the
pulsar, it is given by \cite{1991ApJ...366..501D}:
\begin{equation}
\label{eq:pdotmdot}
\left( \frac{\dot{P}_{\rm b}}{P_{\rm b}} \right)^{\rm \dot{m}} \, = \, \frac{8 \pi G}{T_{\odot} c^5}
\frac{I}{M_{\rm tot}} \frac{\dot{P}_{\rm int}}{P^3} \, \sim \, 3.8 \, \times \, 10^{-21} \rm s^{-1}
\end{equation}
where $T_{\odot} = G M_{\odot} c^{-3} = 4.925490947 \mu \rm s$ is a
solar mass ($M_{\odot}$) in time units, $c$ is the speed of light and
$G$ is Newton's gravitational constant, $I$ is the moment of inertia of the pulsar,
$I \, \simeq \, 10^{38} \rm \, kg\, m^2$. 
Thus $\dot{P}_{\rm b}^{\dot{m}} \, = \, 1.05\, \times \, 10^{-14}\, \rm s\, s^{-1}$,
which is about 40 times smaller than $\dot{P}_{\rm b, kin}$. Finally, the last term
in eq.~\ref{eq:pbdot} is caused by tidal dissipation. 
For PSR\,J2234+0611, this term should be negligible: the WD mass and atmospheric parameters, indicate that the star is well within its Roche lobe and no mass loss occurs. Consequently, the tidal dissipation timescale \citep{1977A&A....57..383Z} is of order 20\,Gyr, well above the characteristic age of the pulsar; $\tau_{\rm c} \simeq 1.5$\,Gyr.

Thus the only relevant term appears to be $\dot{P}_{\rm b, kin}$. This matches the observation
($\dot{P}_{\rm obs} \, = \, 3.3\, \pm \, 2.5 \, \times \, 10^{-12} \, \rm s \, s^{-1}$ for the
DDGR and DDFWHE solutions,
$\dot{P}_{\rm obs} \, = \, 4.9\, \pm \, 2.5 \, \times \, 10^{-12} \, \rm s \, s^{-1}$ for the
DDK solution, see Table~\ref{tab:orbital_solutions})
well; for the DDK solution we have a 2-$\sigma$ ``detection'' of this effect. 

\begin{figure*}
\begin{center}
\includegraphics[width=\textwidth]{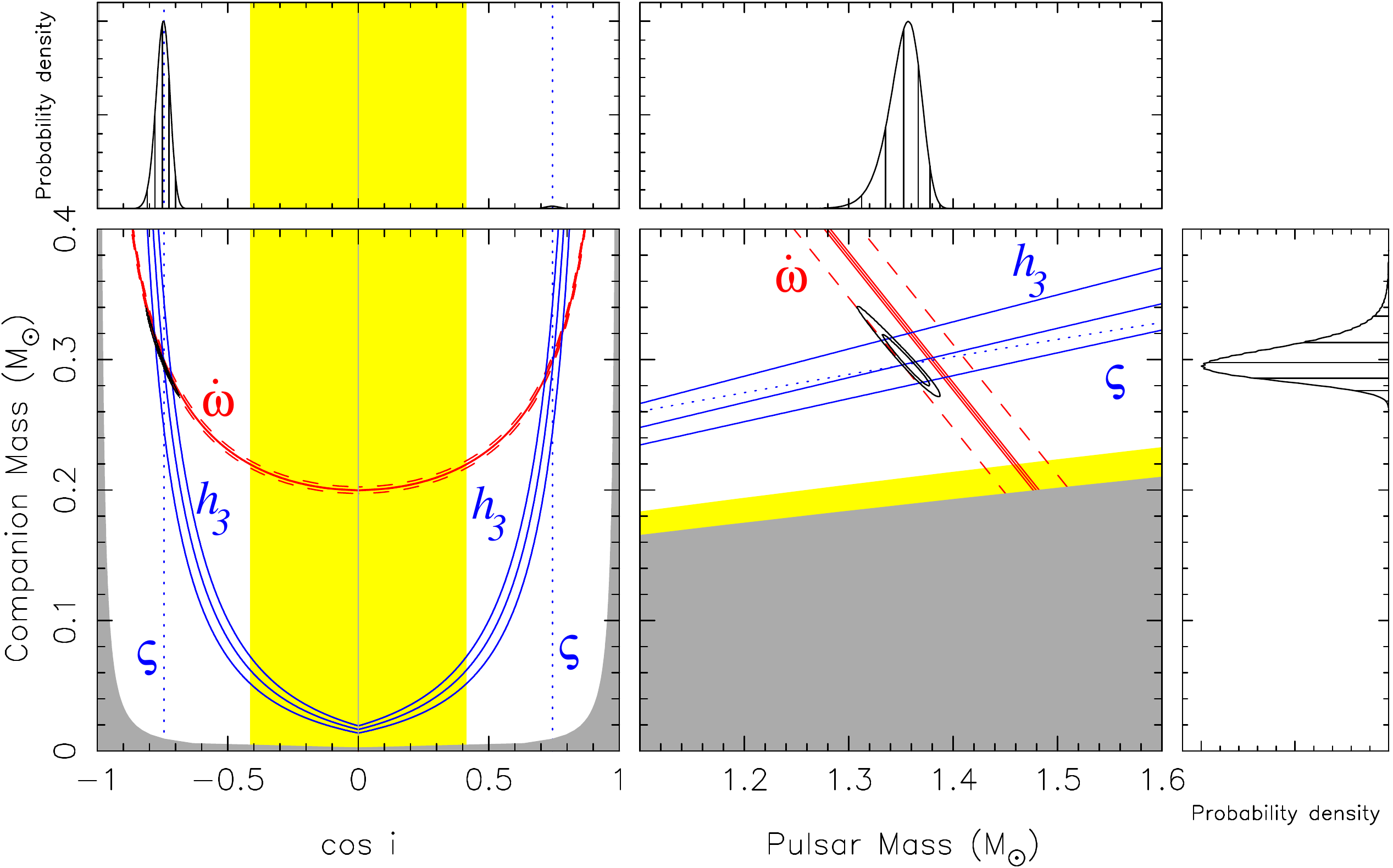}
\caption{Mass constraints for PSR~J2234+0611.
In the main plot on the left we display the $\cos i$ - $M_c$ plane; the gray
region is excluded by knowledge of the mass function and the fact that
the pulsar mass ($M_p$) must be larger than 0.
In the main plot on the right, we display the $M_p$-$M_c$ plane; the gray region
is excluded by knowledge of the mass function and the constraint $\sin i\, \leq\, 1$.
In both plots, the yellow region is excluded by the measurement
of $\dot{x}$. The black contours include 68.23 and 95.44\% of the total
probability density functions (pdf) derived from a 3-D quality
($\chi^2$) map of the $\cos i$-$\Omega$-$M_{\rm tot}$ plane using the DDK orbital model,
with the additional assumption that GR is the correct theory of gravity, see text for details.
The solid blue lines indicate the regions that are (according to GR)
consistent with the nominal and $\pm\, 1-\sigma$ measurements of $h_3$ (solid)
in the DDFWHE model, the blue dashed lines indicate the assumed  $\varsigma$ (dashed)
in that model (see Table~\ref{tab:orbital_solutions}). The solid red lines
indicate the 0, $\pm \,1$-$\sigma$ constraints derived from the
$\dot{\omega}_{\rm obs}$ in the DDFWHE model,
these are equivalent to the $M_{\rm tot}$ in the DDGR model.
The dashed red lines indicate the minimal and maximal values of $M_{\rm tot}$ taking into
account the full range of possible contributions
of the proper motion to $\dot{\omega}_{\rm obs}$,
this is
 $\dot{\omega}_{\rm k} \, = \, \pm \, \mu / \sin i$}
(see text for details). The side panels display the 1-d pdfs
for $\cos i$ (top left), $M_p$ (top right) and $M_c$ (right). The vertical lines
in these pdfs indicate the median and the percentiles corresponding to
1 and 2 $\sigma$ around the median.
\label{fig:mass_mass}
\end{center}
\end{figure*}

\subsection{Post-Keplerian effects. II. Secular rate of advance of periastron}
\label{sec:omdot}

The post-Keplerian effect measured to highest significance for PSR~J2234+0611 is the rate of advance
of periastron, $\dot{\omega}$. According to \cite{2004hpa..book.....L}, the observed effect is given, in the absence of a third component in the system, by:
\begin{equation}
\label{eqn:omdot}
\dot{\omega}_{\rm obs} \, = \, \dot{\omega}_{\rm rel} + \dot{\omega}_{\rm k} + \dot{\omega}_{\rm SO}
\end{equation}
The third term is caused by spin-orbit coupling, a result of the finite size of the
companion white dwarf, for wide systems like PSR~J2234+0611 this effect is negligible.

The first term is caused by relativistic effects. Assuming GR, we can estimate
the total mass of the binary, $M_{\rm tot}$  \citep{rob38} from $\dot{\omega}_{\rm rel}$
by inverting the well-known expression derived by \cite{1982ApJ...253..908T}:
\begin{equation}
\label{eqn:M}
M_{\rm tot}\, = \, \frac{1}{T_{\odot}} \left[ \frac{\dot{\omega}_{\rm Rel}}{3} (1- e^2) \right]^{\frac{3}{2}} \left(\frac{P_{\mathrm{b}}}{2\pi} \right)^{\frac{5}{2}}.
\end{equation}
The DDGR model assumes that $\dot{\omega}_{\rm rel} \, = \, \dot{\omega}_{\rm obs}$, i.e., all
other terms are assumed to be negligible. As we see below this assumption cannot be made
for PSR~J2234+0611. From this assumption, the DDGR model obtains
$M_{\rm tot} \, = \, 1.6798(29) \, \rm M_{\odot}$. The $\dot{\omega}$ provided by the DDFWHE solution
yields, assuming GR, an identical $M_{\rm tot}$. This constraint is
represented by the solid red line in Figure~\ref{fig:mass_mass}.

However, the $\dot{\omega}$ measured by the DDK solution is smaller than that measured by the
DDFWHE model by a small ($\Delta \dot{\omega}\, = \, 9.60 \, \times \, 10^{-6}\, \deg \, \rm yr^{-1}$) but
highly significant (9.3\, $\sigma$) amount. The reason is that, for PSR~J2234+0611,
as for another wide, precisely timed system, PSR~J1903+0327 \citep{2011MNRAS.412.2763F}
the second term in eq.~\ref{eqn:omdot}, $\dot{\omega}_{\rm k}$, is
larger than the measurement uncertainty. This term is given by \cite{1995ApJ...439L...5K},
here re-arranged as in \cite{2011MNRAS.412.2763F}:
\begin{equation}
\label{eq:omdot_k}
\dot{\omega}_{\rm k} \, = \, \frac{\mu}{\sin i} \cos \left( \Theta_{\mu} - \Omega \right),
\end{equation}
where $\Theta_{\mu}$ is the position angle (PA) of the proper motion
and $\Omega$ is the PA for the line of nodes.
In the DDK orbital model, the PAs are measured
in Equatorial (J2000) coordinates, starting from North
through East and an inclination smaller than $90^\circ$ corresponds to a system
where the line-of-sight component of the angular momentum points towards the Earth.

Although $\Theta_{\mu}$ is measured directly from the proper motion (see Table~\ref{table:timsol}),
the orientation of the line of nodes $\Omega$ is generally harder to determine.
In fig.~\ref{fig:mass_mass}, we display with the dashed red lines the total masses
assuming minimal or maximal contributions of $\dot{\omega}_{\rm k}$ to $\dot{\omega}_{\rm obs}$,
$M$
(estimated from equation~\ref{eq:omdot_k} by setting $\cos \left( \Theta_{\mu} - \Omega \right)\, = \, \pm \, 1$).
This shows clearly that $\dot{\omega}_{\rm k}$ is potentially much larger than the
uncertainty in the measurement of $\dot{\omega}_{\rm obs}$.

However, we can estimate the total mass more accurately since, in the
DDK model, we can determine $i$ and $\Omega$ with high precision 
(see details in section~\ref{sec:xdot}, and DDK solution in Table~\ref{tab:orbital_solutions}).
Using these values the model internally estimates
$\dot{\omega}_{\rm k}$ and automatically subtracts it
from the ``measured'' $\dot{\omega}_{\rm obs}$, reporting only the part (presumably)
caused by the relativistic effects, $\dot{\omega}_{\rm rel}$.
Assuming GR, this yields a lower binary mass ($M_{\rm tot} \, = \, 1.6526(29)\, \rm M_{\odot}$)
than estimated by the DDGR model. We consider this to be 
accurate since it takes the proper motion into account.

\subsection{Post-Keplerian effects. III. Shapiro delay}
\label{sec:shapiro}

In the DDGR model, we not only obtain a precise (but in this case innacurate)
estimate for $M_{\rm tot}$, but also
a precise estimate for the companion mass ($M_c \, = \, 0.300(13) \, \rm M_{\odot}$).
Given the mass function of the system, the estimated $M_c$ implies 
$\sin i \, \sim \, 0.668$; this implies either $i \, \sim 42 \deg$
or $i \, \sim 138 \deg$. The measurement is
possible because of the presence of the Shapiro delay, however, the fact that
the Shapiro delay is detected at all for 
a system with an orbital inclination so far from edge-on (90$^\circ$)
is unusual. The detection in this case is the is result of two factors: one is the high
timing precision of the
Arecibo observations of this pulsar, the second is the large eccentricity
of the orbit; the latter helps separate the Shapiro delay from the normal ``Roemer''
delays caused by the geometry of the orbital motion relative to the line of sight.

\begin{figure*}
\begin{center}
\includegraphics[width=0.75\textwidth]{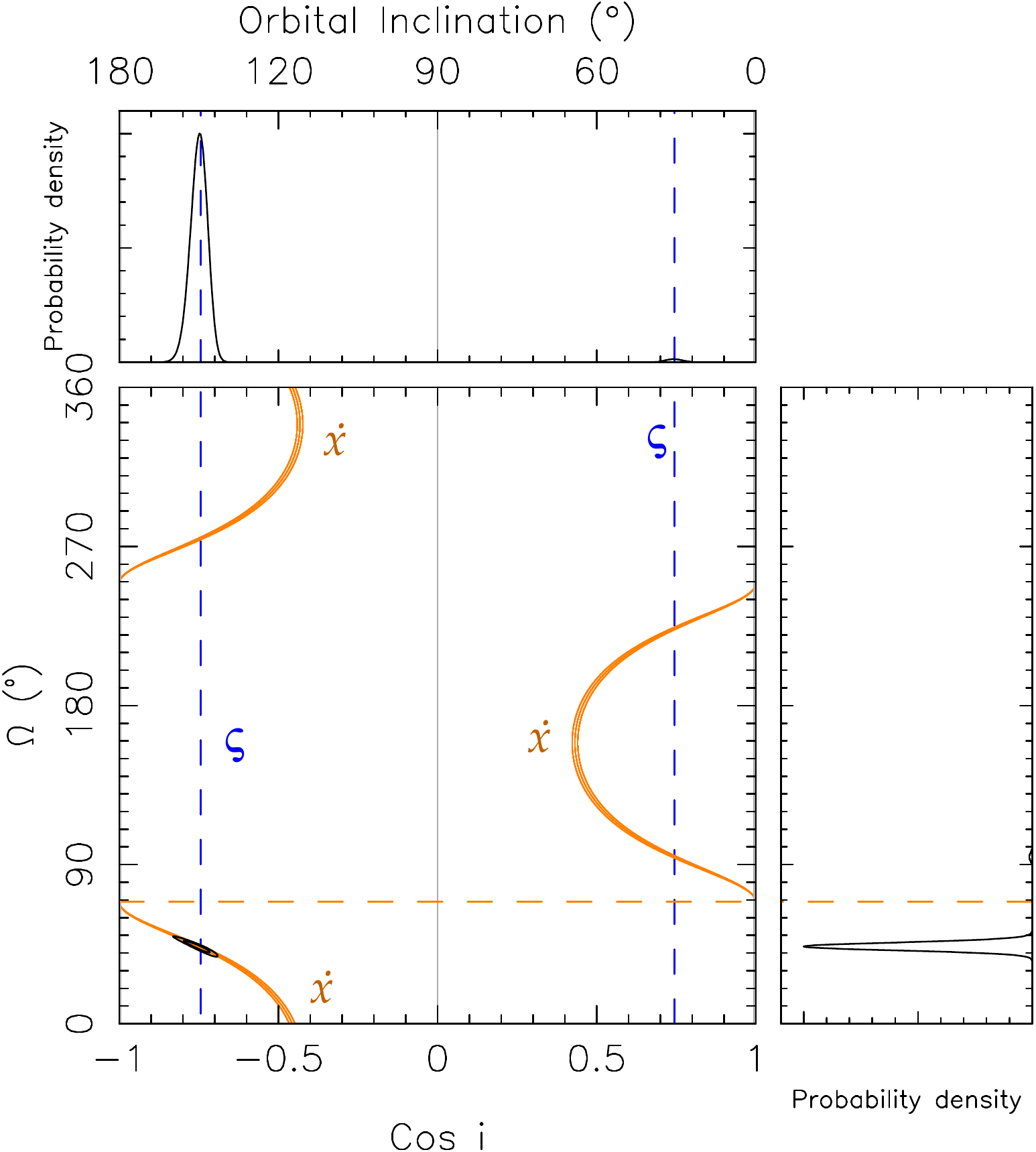}
\caption{
Orbital orientation constraints for PSR~J2234+0611. In the main square
panel we display the full $\cos i$-$\Omega$ plane; this
has {\em a priori} a constant probability density for randomly aligned systems.
The black contours include 68.23 and 95.44\% of the total
probability density function (pdf) derived from a 3-dimensional $\chi^2$ map of
$\cos i$-$\Omega$-$M_{\rm tot}$ space using the DDK model with the additional
assumption that GR is the correct theory of gravity.
The dashed orange line indicates the PA of the proper motion of the system
($\Theta_{\mu}\, = \, 69.0(1) \, \deg$).
The dashed blue line indicates the $\varsigma$ assumed in
our DDFWHE model; the solid orange lines indicate the regions that are 
consistent with the nominal and $\pm\, 1-\sigma$ measurements of
$\dot{x}$ obtained in that model (see Table~\ref{tab:orbital_solutions}).
The $\varsigma$ and $\dot{x}$ constraints predict well the location of the
region(s) of high probability but
provide no distinction between the four locations where they cross
(these are listed in Table~\ref{tab:grid}); this can
only be done using the DDK model (see text for details).
The side panels display the 1-d pdfs for $\cos i$ (top) and $\Omega$ (right).}
\label{fig:cosi_omega}
\end{center}
\end{figure*}

The far from edge-on inclination means that the Shapiro delay is not 
easy to measure. When using the DDFWHE model to fit for
both Shapiro delay parameters, $h_3$ and $\varsigma$, both values 
are measured to relatively low confidence.
In order to better quantify the Shapiro delay, we use the best-fit value of $s \equiv \sin i$
that corresponds to the masses determined by the DDGR model ($s \, = \, 0.6677654...$) to
derive \citep{2010MNRAS.409..199F}:
\begin{equation}
\varsigma \, = \, \frac{s}{1 + \sqrt{1 - s^2}} = 0.382811...,
\end{equation}
this is represented by the blue dashed line in figures~\ref{fig:mass_mass} and
\ref{fig:cosi_omega}. Fixing this in the DDFWHE model, we obtain a significant
$h_3 \, = \, 82 \, \pm 14\, \rm n s$; an unusually small value that is a consequence 
of the inclination of the system. The mass and inclination constraints 
introduced by this measurement and its $\pm 1$-$\sigma$ uncertainties are shown
by the solid blue curves in figure~\ref{fig:mass_mass}. 
The region where these $h_3$ lines cross
the $\dot{\omega}$ lines provides a good explanation of the DDGR
estimate for $M_c$ and its related uncertainty.

\subsection{Secular change of the projected semimajor axis}
\label{sec:xdot}

As seen in Table~\ref{tab:orbital_solutions}, both the DDGR and DDFWHE timing solutions
contain a precise measurement of a change in the projected semi-major axis ($\mathrm{x = a \sin i}$ where a is the
semi-major axis and i is the orbital inclination) of the
pulsar's orbit, $\dot x \, =\, -2.79(7)\times10^{-14}\, \rm lt-s\, s^{-1}$,
thus $(\dot x / x)^{\rm obs} \, =\, -1.99(5)\times10^{-15}\, \rm s^{-1}$.
Following  \cite{2004hpa..book.....L}, the observed change in $\dot x$ can,
in the absence of a third object in the system, be written
in terms of various contributions as:
\begin{equation}
\label{eq:xdot}
\left( \frac{\dot x}{x} \right)^{\rm obs}\,  = \,
\left( \frac{\dot x}{x} \right)^{\rm k} \, + \,
\left( \frac{\dot x}{x} \right)^{\rm GW} \, + \,
\frac{d \epsilon_A}{dt} \, - \,
\frac{\dot{D}}{D} \, + \,
\left( \frac{\dot x}{x} \right)^{\dot{m}} \, + \,
\left( \frac{\dot x}{x} \right)^{\rm SO}.
\end{equation}
The first term is caused by the changing geometry due to
the motion of the system relative to the Earth, it is given by \citep{1995ApJ...439L...5K}:
\begin{equation}
\label{eq:xdotk}
\dot x_{\rm k}  = x \mu \cot i \sin ( \Theta_\mu - \Omega),
\end{equation}
where we have, again, re-written the terms as in \cite{2011MNRAS.412.2763F}, except for the latter's
negative sign, so that we are in the right-handed convention being used 
in the DDK model.
As we will see below, this is the only term that can account
for the observations.

The second term is from the decrease of the size of the orbit caused by gravitational wave emission;
this is given by
\begin{equation}
\label{eq:adot}
\left( \frac{\dot{x}}{x} \right)^{\rm GW} \, = \, \frac{2}{3} \frac{\dot{P}_{\rm b, GW}}{P_{\rm b}} \, = \, -6.31 \times \, 10^{-24} \, \rm  s^{-1},
\end{equation}
where we used the predicted $\dot{P}_{\rm b, GW}$ from the DDGR solution.
This is more than eight orders of magnitude smaller than the measured value.

The third term, caused by aberration, is proportional to the geodetic precession rate for
the pulsar, this is given by \cite{1975PhRvD..12..329B} as:
\begin{equation}
\Omega_{\rm geod} \, = \, \left( \frac{2 \pi}{P_{\rm b}} \right)^{5/3}
T_{\odot}^{2/3} \frac{1}{1 - e^2} \frac{M_c (4 M_{\rm tot} - M_c) }{2 M_{\rm tot}^{4/3}},
\end{equation}
the result for PSR~J2234+0611 is
$\Omega^{\rm geod} \, = \, 5.6 \, \times \, 10^{-14} \, \rm rad \, s^{-1}$, i.e., the
geodetic precession cycle has a length of 3.6 Myr. The aberration contribution to $\dot{x}$ is then given by 
\citep{1992PhRvD..45.1840D}:
\begin{equation}
\frac{d \epsilon_A}{dt}\, =\, \frac{P}{P_b} \frac{\Omega^{\rm geod}}{\sqrt{1 - e^2}}
\frac{\cot \lambda \sin 2 \eta + \cot i \cos \eta}{\sin \lambda},
\end{equation}
where $\eta$ and $\lambda$ are the polar coordinates of the pulsar's spin. For PSR~J2234+0611,
the non-geometric factors (the first two fractions in the equation above)
amount to $-7.28 \, \times \, 10^{-23} \rm \, s^{-1}$, making this
term about 8 orders of magnitude smaller than the observed value.

The fourth term, $-\dot{D} / D \, = \, 1.541\, \times\, 10^{-18}\, s^{-1}$, is three orders of
magnitude smaller than the observed effect.

The fifth term can be derived from eq.~\ref{eq:adot}, with the orbital variability
given by eq.~\ref{eq:pdotmdot}, from this we obtain
$(\dot{x} / x)^{\dot{m}}\, = \, 2.53 \, \times \, 10^{-21} \rm \, s^{-1}$, which is also
extremely small.

Finally, the sixth term is due to changes in the orbital plane of the system from spin-orbit coupling; these are extremely small in such a wide system.

Since the first term in eq.~\ref{eq:xdot} is the only measurable contribution to
$\dot{x}_{\rm obs}$, we will now assume that the latter is described by eq.~\ref{eq:xdotk}.
Using that equation, we can combine $\dot{x}_{\rm obs}$ with the Shapiro delay to constrain the system
geometry as shown in Figure~\ref{fig:cosi_omega}. The orange lines show the
$\cos i$ and $\Omega$ that are consistent with the measured $\dot x$ while the dashed blue
lines show the $\cos i$ compatible with the assumed $\varsigma$.
These cross in four locations, listed in Table~\ref{tab:grid}; these represent
the four possible orbital orientations of the system according to
the DDK and DDFWHE timing solutions.

\subsection{Annual orbital parallax}
\label{sec:aop}

For most pulsars where these constraints are available, we cannot eliminate the degeneracy
implied by these four $i$-$\Omega$ solutions. However, if the binary system is relatively
nearby and has a high timing precision, then apart from the secular variation of $\omega$
($\dot{\omega}_{\rm obs}$) and $x$ ($\dot{x}_{\rm obs}$) there are yearly cyclical
variations in these parameters caused by Earth's orbit around the Sun \citep{1996ApJ...467L..93K}.
These are taken into account in the DDK model.

In Table~\ref{tab:grid}, we can see that the quality of the local
$\chi^2$ minima are clearly not identical, being significantly better
for solution 1 (this is the DDK solution presented in Table~\ref{tab:orbital_solutions});
the latter solution is significantly better than either the DDGR or the DDFWHE solutions.
A possible reason for this is that we have detected the yearly cyclical variations of
$x$ or $\omega$ or both. We quantify this statement in the next section.

\section{Bayesian analysis of the system}
\label{sec:bayesian}

Before we proceed, we emphasize that no single orbital model captures all features of the
system in a self-consistent way. The DDGR and DDFWHE models over-estimate
$M_{\rm tot}$ and $\dot{\omega}_{\rm rel}$ respectively
(and because of that $M_{p}$ and $M_c$ as well) because they do not take into account
$\dot{\omega}_K$. The DDK model captures the kinematic effects well and provides
an accurate estimate of $\dot{\omega}_{\rm rel}$, but it has
a larger than necessary uncertainty on $M_c$ (about $0.05\, M_{\odot}$,
even with a fixed orbital inclination)
because it uses a sub-optimal parameterization of the Shapiro delay and does not
assume the validity of GR.

\subsection{Mapping the $\Omega$-$\cos i$-$M_c$ space}

Given all the correlations and caveats related to the different orbital models,
and in order to better determine $M_{\rm tot}$, $M_c$, $M_p$, $i$, $\Omega$,
their uncertainties and correlations, we have
made a self-consistent $\chi^2$ map of the $\Omega$-$\cos i$-$M_{\rm tot}$
space using the DDK orbital solution with the assumption that GR is the
correct theory of gravity. These parameters are chosen because they have
{\it a priori} a constant probability density for randomly aligned orbits.

For each point in the grid of $i$, $\Omega$, and $M_{\rm tot}$
values, we hold $\Omega$ and $i$ fixed in the DDK model (from this it estimates
all kinematic effects) and derive other relevant
post-Keplerian parameters from the known mass function, $i$ and $M_{\rm tot}$
(M2; OMDOT and GAMMA)
using the GR equations. All these parameters are fixed inputs to the
DDK model used to do the timing analysis for that grid point.
The Einstein delay (GAMMA) must be calculated and used in the fit because, for wide binary
pulsars like PSR~J2234+0611, it is strongly correlated with $\dot{x}$ in the DDFWHE model
and with $\Omega$ in the DDK model (see Ridolfi et al. 2018, in preparation).
We then run {\tt tempo}, fitting for all other timing parameters not mentioned above,
recording the value of $\chi^2$ for each combination of $\Omega$, $\cos i$ and $M_{\rm tot}$.

Given the computational expense, our $\chi^2$ map does not cover the 
full space; it consists instead of four disconnected regions around
the four local $\chi^2$ minima listed in Table~\ref{tab:grid}; the $\cos i$
and $\Omega$ bounds sampled around these minima are also listed there.
These variables are sampled with step sizes of 0.004 and $0.2\, \deg$, respectively.
 For each grid section, we mapped the third variable, $M_{tot}$ from 1.641$M_\odot$ to
1.731$M_\odot$ with a step size of 0.0006$M_\odot$. 
The quality of the fits in the regions outside these bounds are
extremely low, for that reason those regions were not sampled.

The resulting 3-D grids of $\chi^2$ values are then used to calculate a 3-dimensional
probability density function (pdf) for $\Omega, \cos i, M_{\rm tot}$, as
discussed by \cite{2002ApJ...581..509S}:
\begin{equation}
p(\Omega, \cos i, M_{\rm tot}) \propto e^{(\chi^2_{\rm min} - \chi^2)/2},
\end{equation}
where $\chi^2_{\rm min}$ is the lowest $\chi^2$ of the whole grid.

This 3-D pdf is then projected onto two planes:
the $\cos i - M_c$ plane and derived $M_c$-$M_p$ plane (see contours in the main panels
in figure~\ref{fig:mass_mass}) and the $\cos i$-$\Omega$ plane
(see contours in main panel of figure~\ref{fig:cosi_omega}).
It is also projected along three axes, $\cos i$ (top left side panels
in figures~\ref{fig:mass_mass} and \ref{fig:cosi_omega}), $M_c$ (and derived $M_p$,
see top right and right side panels in figure~\ref{fig:mass_mass})
and $\Omega$ (depicted in the right side panel of figure~\ref{fig:cosi_omega}).

\subsection{Results of the Bayesian analysis}

The resulting 1-D pdfs show that, as hinted by the $\chi^2$ values in
Table~\ref{tab:grid}, the probabilities for the four $i$-$\Omega$ solutions
are far from identical. The solution with the lowest $\chi^2$
(number 1) is preferred, with a total probability of 98.786\%. The second most
likely solution (number 2), has a total probability of 1.214\%, it is
still visible in the side panels of Fig.~\ref{fig:cosi_omega} as a separate
peak with very small amplitude. Solutions 3 and 4 have
probabilities that are too small for our numerical precision, being thus
definitively excluded. The discrimination between solutions 1 and 2, i.e., between
the two possible values of $i$ and $\Omega$ does not yet reach a statistical significance
equivalent to 3$\sigma$, but they imply that the absolute orbital orientation of
the system will be precisely known in the near future.
Despite the fact that we cannot yet point
out a single solution to equivalent 3-$\sigma$ significance, the exclusion of two of the
solutions to high significance implies a significant detection of the annual orbital
parallax.

The values derived for the quantities we set out to determine are:
$i = 138.7^{+2.5}_{-2.2}\, \deg$ (68.27 \% C. L.),
$138.7^{+5.1}_{-4.2} \deg$ (95.45 \% C. L.),
$\Omega \, = \, 43.7^{+2.3}_{-2.2} \deg$ (68.27 \% C. L.) and
$43.7^{+5.2}_{-4.4} \deg$ (95.45 \% C. L.). In Fig.~\ref{fig:cosi_omega}, we
see a fine correlation between $i$ and $\Omega$, which is a direct consequence
of the precisely measured $\dot{x}$.

For the component masses, the situation is very clear: both solutions with
measurable probability have the same total mass,
$M_{\rm tot} \, = \, 1.6518^{+0.0033}_{-0.0035} \, \rm M_{\odot}$ (68.27 \% C. L.),
$1.6518^{+0.0066}_{-0.0070} \, \rm M_{\odot}$ (95.45 \% C. L.).
For the component masses we obtain
$M_c \, = \, 0.298^{+0.015}_{-0.012}\, \rm M_{\odot}$ (68.27 \% C. L.),
$0.298^{+0.034}_{-0.021}\, \rm M_{\odot}$ (95.45 \% C. L.),
$M_p \, = \, 1.353^{+0.014}_{-0.017}\, \rm M_{\odot}$ (68.27 \% C. L.) and
$1.353^{+0.025}_{-0.040}\, \rm  M_{\odot}$ (95.45 \% C. L.).
The fine $\Omega$-$i$ correlation also affects the mass
measurements: for values of $\Omega$ closer to $\Theta_\mu$, the more
face-on inclinations result in a more massive companion and a less massive
pulsar.

The measurements made by the Bayesian analysis are in good 
agreement with the values inferred by the results in 
Section~\ref{sec:results}. For example, the total mass is
well described by
the $\dot{\omega}$ of the DDK solution, as it must since we used the
latter model to map the masses assuming GR.
The individual masses are well described by the intersection of
the latter constraint with the $h_3$ of the DDFWHE solution. The constraints
these impose on the range of inclinations plus the constraints imposed 
by the detection of $\dot{x}$ in the DDGR/DDFWHE solutions
provide a good description of the range of $\Omega$, plus its strong
correlation with $i$ near the best DDK solution.

\section{Implications}
\label{sec:discussion}

The mass of PSR~J2234+0611 is very similar to that of PSR~J1807$-$2500B
in the globular cluster NGC~6544 ($M_p \, = \, 1.3655(21) \, M_{\odot}$, \citealt{2012ApJ...745..109L}),
PSR~J1713+0737 ($M_p \, = \, 1.33^{+0.09}_{-0.08} \, M_{\odot}$, 
\citealt{2018ApJS..235...37A} or $M_p \, = \, 1.35(7) \, M_{\odot}$,
\citealt{2016MNRAS.458.3341D}). Until recently these would have been
considered unusually small
masses for a fully recycled pulsar. However, recent measurements show
that two other fully recycled pulsars might be even less massive:
PSR~J1918$-$0642 ($M_p \, = \, 1.29^{+0.10}_{-0.09} \, M_{\odot}$
\citealt{2018ApJS..235...37A}) and PSR~J0514$-$4002A, a 5-ms pulsar
located in the globular cluster NGC~1851
($M_p \, = \, 1.25^{+0.06}_{-0.05} \, M_{\odot}$, see Ridolfi et al. 2018,
in preparation).

Such low masses are interesting because they can constrain
the efficiency of the recycling 
process \citep[for a detailed discussion see][; an update of that discussion is presented by Ridolfi et al., in prep]{2012MNRAS.423.3316A}.
If PSR~J2234+0611 indeed descended from a typical LMXB (section~\ref{sec:formation}), then the system parameters reported here imply,
following the arguments presented by \cite{2016ApJ...830...36A},
a mass-accretion efficiency (the fraction of mass lost by the donor that is accreted onto the neutron star) of at most $\sim30\%$ for an initial pulsar mass $\le 1.17$\,M$_{\odot}$, or $\sim6\%$ for a more typical initial mass of 1.35\,M$_{\odot}$.

\subsection{Formation of eccentric MSPs}
\label{sec:formation}

Our analysis of PSR J2234$+$0611  is informed by previous work on
two other eccentric binary systems, PSRs J1946$+$3417 and 
J1950$+$2414. Mass measurements of these three pulsars can be
combined to constrain theories of formation for eccentric binary
MSPs.
The rotation-delayed accretion induced collapse (RD-AIC) hypothesis presented
by \cite{2014MNRAS.438L..86F} for the formation of the eccentric MSPs
has been excluded already by the mass
measurement for PSR~J1946+3417 presented in \cite{2017MNRAS.465.1711B}:
the mass of that pulsar ($M_p \, = \, 1.828(22) \, M_{\odot}$) is
too large to have resulted from the collapse of a massive WD.

The RD-AIC theory could in principle generate larger MSP masses
if we allow for differential rotation of the massive WD progenitor
to the MSP: With differential rotation WDs can be
much more massive than the $\sim \, 1.48\, M_{\odot}$ upper
mass limit for rigidly rotating WDs.
However, even in such a case the systems produced by RD-AIC would still
have small peculiar velocities, otherwise the range of observed
orbital eccentricities wouldn't be as small as the observed range
(see details in \citealt{2014MNRAS.438L..86F}). Such a possibility
is difficult to reconcile with the large peculiar velocity measured for
PSR~J1946+3417 (in particular its large vertical velocity relative to the Galaxy,
see \citealt{2017MNRAS.465.1711B}) and the observed velocity of
PSR~J2234+0611 \citep{2016ApJ...830...36A}.

The large mass of PSR~J1946+3417 is consistent with the hypothesis proposed
by \cite{2015ApJ...807...41J}, which is also based on an instantaneous
loss of binding energy of the more massive component. However,
in this hypothesis the more massive component starts as a
massive MSP. As it spins down, the centrifugal support is steadily reduced,
causing a slow but steady increase in the central pressure with time,
until a critical threshold is reached and the phase transition
happens, presumably forming a quark star or some other type of 
exotic object that is still observable as a MSP.
The sudden decrease in mass (owing to the larger binding energy of
the new exotic remnant) results in the large orbital eccentricity.
Other properties of PSR~J1946+3417, like the large vertical velocity
relative to the Galactic disk, are also consistent with this hypothesis, since
the original system already shared the large velocity (relative to typical
stars in the Galactic disk) typical of MSP-WD binaries.
However, if there is a single pressure threshold for this phase transition,
the masses observed  for the MSPs in these eccentric systems should lie in a relatively
narrow range (which is nevertheless finite because of
differences in the spin periods, which would result in 
different NS masses for the same central pressure at which the 
phase transition occurs).

The masses measured for PSR~J2234+0611 ($1.35 \, \rm M_{\odot}$)
and for PSR~J1950+2414 ($M_p \, = \, 1.495(24)\, M_{\odot}$, Zhu et al. 2018, in preparation)
are inconsistent with this hypothesis, since they are much smaller
than the mass observed for PSR~J1946+3417 - clearly,
a single pressure threshold for a phase transition does not provide a good description of
these systems. Indeed, the observed MSP masses within this class
appear to be as broad as observed for the general MSP population
\citep{2016ARA&A..54..401O,2016arXiv160501665A}. 

All measurements thus far are consistent with the expectations of the
hypothesis proposed by \cite{2014ApJ...797L..24A}. This
proposes that the orbital eccentricity is caused by
material ejected from the companion due to unstable hydrogen shell burning.
This hypothesis predicts that the
MSPs in these systems should have a range of masses and
Galactic velocities similar to those of the general MSP population;
the observations are thus far consistent with this prediction.

Regarding the companions to the MSPs in these systems, all hypotheses
advanced to date predict that they should be Helium
white dwarfs with masses similar to what should be expected
from the \cite{1999A&A...350..928T} relation. For PSRs 
J1946$+$3417, J1950$+$2414, and J2234$+$0611 the mass ranges
predicted by this relation are 0.275--0.303, 0.268--0.296,
and 0.281--0.310, respectively.
In the case of PSR~J2234+0611,
our measured WD mass is in agreement with that prediction. For PSR~J1946+3417,
the companion mass 
$M_c \, = \, 0.2556(19)\, M_{\odot}$ is marginally consistent with
this expectation, being lighter than expected. The companion of PSR~J1950+2414
has a mass ($M_c \, = \, 0.280_{-0.004}^{+0.006}\, M_{\odot}$) that is also
well within the range expected by \cite{1999A&A...350..928T} for its orbital period. 
We note that within the context of the CB disk scenario, depending on the lifetime and mass of the disk, there could be a significant (up to $\sim 10\%$) reduction of the orbital separation. This effect would result in somewhat larger masses for a given orbital period, compared to the \cite{1999A&A...350..928T} relation ---  the opposite of what is observed for  PSR~J1946+3417.

\subsection{White dwarf properties}
The distance to PSR J2234+0611 is very well measured through the detection of timing parallax,
$\mathrm{\varpi=1.03(4)}\, \rm mas$. This corresponds to a distance $d \, =\, 0.97(4)$ kpc. The
uncertainty of 40 pc for J2234+0611 places it among the best measured pulsar distances.

The distance estimate will improve further in the near future. As the timing
baseline $T$ increases, the precision of $\dot{P}_{\rm b, obs}$ will also improve quickly.
This will result in an additional precise distance estimate from the inversion of eq.~\ref{eq:pbdot}, which
will only be limited by knowledge of the Galactic potential. Measurements of
this distance can be corroborated by VLBI campaigns.
The component masses will also improve significantly, particularly the total mass;
for the individual masses significant improvements will depend on advances in
timing precision.

The precise distance and mass estimates presented here, together with the spectroscopic constraints on the WD atmospheric properties, transform the system into a laboratory for testing WD physics. As discussed in detail in \cite{2016ApJ...830...36A}, the aforementioned measurements yield a radius estimate of $R_{\rm WD}=0.024_{-0.002}
^{+0.004}$\,R$_{\odot}$ and a surface gravity of $\log g = 7.11_{-0.16}^{0.08}$\,dex, both of which are model-independent. This is important for two reasons: firstly, 
PSR\,J2234+0611 is only the second system after PSR\,J1909$-$3744 for which independent atmospheric parameters can be obtained \citep{2016ApJ...830...36A}. Second, the surface 
temperature of $T_{\rm eff}\simeq 8600$\,K obtained from atmospheric modeling, indicate that the WD envelope is convective. 
Spectroscopic 1D models for cool convective 
atmospheres are  suspected to produce spurious results, but quantitative estimates and empirical corrections are difficult to obtain due to the lack of measurements. For both 
these reasons, PSR\,J2234+0611 becomes particularly important for calibrating atmospheric models. Currently, the precision of such tests is severely limited by the poor quality 
of the optical spectra, but could be improved significantly with further optical observations.  

In addition, PSR\,J2234+0611 can also be used to test the predictions of WD mass-radius relations. One of the main remaining uncertainties in low-mass WD cooling models is the 
size of the hydrogen envelope that surrounds the degenerate Helium core. The latter can significantly affect the stellar radius, as well as the main energy source (residual 
hydrogen shell burning vs thermal cooling) and, consequently, the cooling age. 
Here again, our estimates are broadly consistent with the predictions for thin-envelope models, but a detailed test is limited by measurement uncertainties of the WD atmospheric parameters \citep[see][for details]{2016ApJ...830...36A}. For PSR\,J2234+0611, a future precision measurement of its envelope size is also important for probing its formation history, since the thin-shell instabilities on the proto-WD required for creating a CB disk, are also expected to reduce significantly the size of the WD envelope \citep[see][and references therein]{2014A&A...571L...3I,2016A&A...595A..35I,2016ApJ...830...36A}. 

Last but not least, PSR\,J2234+0611 is within a few 100\,K from the ZZ-Ceti instability strip for low-mass WDs. \cite{2018MNRAS.479.1267K} recently reported on photometric observations of the system and found no pulsations. Consequently, the improved mass estimate reported in this work can further constrain the instability mechanism and the structure of WD envelopes \citep[see Figure\, 5 in ][and references therein for details]{2018MNRAS.479.1267K}.   

\section{Conclusions}\label{sec:conc}

We have reported the timing solution for PSR J2234$+$0611, a 3.6-ms pulsar in an
eccentric (e = 0.13), 32-day orbit with a He white dwarf. The pulsar is bright
(especially with Arecibo) and has a narrow pulse and therefore has excellent timing
precision. It was added to pulsar timing array efforts soon after discovery and
therefore is observed regularly. The exceptional timing properties of
this pulsar, its eccentric orbit, and the optical detection has allowed the
precise measurement of an unprecedented number of parameters, indeed,
this is the first binary pulsar where we know the precise 3-D location and
3-D velocity, the full 3-D orientation of the orbit and, on top of that, we are able to
determine precise masses. To our knowledge, no other binary pulsar has such precisely
determined overall geometry.

We have compared the characteristics of this pulsar to
those expected from various theories for the eccentric MSP systems and show that
the only viable remaining theory is one where mass-loss occurs due to unstable shell-hydrogen burning in the proto-WD \citep{2014A&A...571L...3I,2015ASSP...40....1A,2016A&A...595A..35I}. We expect that this MSP system
will be useful for constraining white dwarf models, given its well measured distance,
white dwarf mass, and optically detectable companion.

\section*{Acknowledgments}
The Arecibo Observatory is operated by the University of Central Florida,
Ana G. M\'{e}ndez-Universidad Metropolitana, and Yang Enterprises under a cooperative
agreement  with the National Science Foundation (NSF; AST-1744119).
The National Radio Astronomy Observatory is a facility of the National Science Foundation operated under cooperative agreement by Associated Universities, Inc. This work was supported
by the NANOGrav Physics Frontiers Center (NSF award 1430284).
P.C.C.F. gratefully acknowledges financial support by the European Research Council,
under the European Union's Seventh Framework Programme (FP/2007-2013) grant agreement
279702 (BEACON) and continuing support from the Max Planck Society. J.S.D. is supported
by the NASA Fermi program. Pulsar research at UBC is supported by an NSERC Discovery Grant
and by the Canadian Institute for Advanced Research. J.G.M. was supported for this research
through a stipend from the International Max Planck Research School (IMPRS) for Astronomy
and Astrophysics at the Universities of Bonn and Cologne. Finally, we thank Norbert Wex for 
the useful suggestions.

\facility{Arecibo}
\software{PRESTO \citep{2001PhDT.......123R,2011ascl.soft07017R},
          Tempo \citep{tempo2015}, PSRCHIVE \citep{Hotan2004}}

\bibliography{ao327_2234}
\bibliographystyle{yahapj}

\end{document}